\documentclass[11pt,twoside]{article}

\usepackage[usenames,x11names,table]{xcolor}
\usepackage{float}
\usepackage[font={rm,md,sl},format=hang]{subfig}
\usepackage[font={rm,small,sl}]{caption}
\usepackage{graphicx}
\usepackage[pagebackref,breaklinks]{hyperref}
\usepackage{breakurl}
\usepackage[margin=1.1in]{geometry}

\hypersetup{%
	pdfusetitle=true,%
	bookmarks=true,%
	pdftitle={Main Memory Adaptive Indexing for Multi-core Systems},%
    pdfauthor={Victor Alvarez, Felix Martin Schuhknecht, Jens Dittrich, Stefan Richter},     
    pdfsubject={Adaptive Indexing},   
    pdfcreator={Victor Alvarez, Felix Martin Schuhknecht, Jens Dittrich, Stefan Richter},   
    pdfproducer={\LaTeX}, 
	pdfkeywords={Information Systems, Adaptive Indexing, Sorting Algorithms, Parallel Algorithms},%
	pdfdisplaydoctitle=true,%
	pdfstartview=Fit,%
	colorlinks=false%
}

\renewcommand*{\backrefalt}[4]{%
\ifcase #1 %
No citations.%
\or
One citation on page #2.%
\else
#3 citations on pages #2.%
\fi
}

\newcommand{\num}[1]{\oldstylenums{#1}}
\newcommand{\algo}[1]{{\textbf{#1}}}

\newcommand{\bencha}{{\tt{Total}}}
\newcommand{\benchb}{{\tt{Initial}}}

\newcommand{\factor}{{\tt{Factor}}}
\newcommand{\algorithm}{{\tt{Algorithm}}}
\newcommand{\moniker}{{\tt{Short name}}}
\newcommand{\threads}{{\tt{Threads}}}
\newcommand{\speedup}{{\tt{Speedup}}}
\newcommand{\slowdown}{{\tt{Slowdown}}}
\newcommand{\SC}{{\textbf{SC}}}
\newcommand{\CGI}{{\textbf{CGI}}}
\newcommand{\RS}{{\textbf{RS}}}
\newcommand{\STL}{{\textbf{STL-S}}}
\newcommand{\PSC}{{\textbf{P-SC}}}
\newcommand{\PCSC}{{\textbf{P-CSC}}}
\newcommand{\PCGI}{{\textbf{P-CGI}}}
\newcommand{\PCCGI}{{\textbf{P-CCGI}}}
\newcommand{\PRS}{{\textbf{P-RS}}}
\newcommand{\PRPRS}{{\textbf{P-RPRS}}}
\newcommand{\PCRS}{{\textbf{P-CRS}}}

\newcommand{\HCS}{{\textbf{HCS}}}

\colorlet{color}{cyan!50}

\begin{document}


\title{Main Memory Adaptive Indexing for Multi-core Systems}


\author{Victor Alvarez\qquad Felix Martin Schuhknecht\qquad Jens Dittrich\qquad Stefan Richter\\
\\
Information Systems Group\\
Saarland University\\
\url{http://infosys.cs.uni-saarland.de}}

\date{\today}

\maketitle

\begin{abstract}

Adaptive indexing is a concept that considers index creation in databases as a by-product of query processing; as opposed to traditional full index creation where the indexing effort is performed up front before answering any queries. Adaptive indexing has received a considerable amount of attention, and several algorithms have been proposed over the past few years; including a recent experimental study comparing a large number of existing methods. Until now, however, most adaptive indexing algorithms have been designed single-threaded, yet with multi-core systems already well established, the idea of designing parallel algorithms for adaptive indexing is very natural. In this regard, and to the best of our knowledge, only one parallel algorithm for adaptive indexing has recently appeared in the literature: The parallel version of standard cracking. In this paper we describe three alternative parallel algorithms for adaptive indexing, including a second variant of a parallel standard cracking algorithm. Additionally, we describe a hybrid parallel sorting algorithm, and a NUMA-aware method based on sorting. We then thoroughly compare all these algorithms experimentally; along a variant of a recently published parallel version of radix sort --- which was observed to be very fast in practice. Parallel sorting algorithms serve as a realistic baseline for multi-threaded adaptive indexing techniques. In total we experimentally compare seven parallel algorithms. Additionally, we extensively profile all considered algorithms w.r.t.~bandwidth utilization, cache misses, NUMA effects and locking times to discover their weaknesses. The initial set of experiments considered in this paper indicates that our parallel algorithms significantly improve over previously known ones. Our results also suggest that, although adaptive indexing algorithms are a good design choice in single-threaded environments, the rules change considerably in the parallel case. That is, in future highly-parallel environments, sorting algorithms could be serious alternatives to adaptive indexing.

\end{abstract}

\section{Introduction}\label{sec:intro}

Traditionally, retrieving data from a table in a database is improved by the use of indexes when highly selective queries are performed. Among the most popular data structures that are used as indexes we can find self-balancing trees, \mbox{B-trees}, hash maps, and bitmaps. However, on the one hand, building an index requires extra space, but perhaps most importantly, it requires time. Without any knowledge about the workload, the best hint a database has to build an index, is to simply build it up front the first time data is accessed. This, ironically, could slow down the whole workload of a database. Furthermore, it could be the case that only few queries are performed over different attributes, or it could simply be the case that response time per query, including the very first one, is the most important measure. In these cases, the up-front effort in building indexes does not really pay off. On the other hand, no index at all is in general not a solution either, as full table scans incur very quickly in high total execution times. Hence, ideally, we would like to have a method that allows us to answer \emph{all} queries as fast as if there were an index, but with initial response time as if there were no index. This is the spirit of adaptive indexing.

\begin{figure}[!htb]
	\begin{center}
		\includegraphics[scale=0.55]{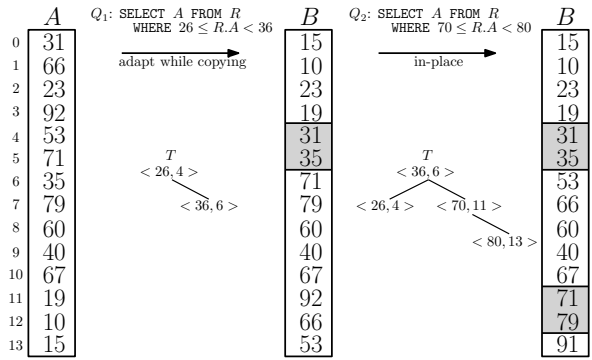}
	\end{center}
	\caption{Adaptively indexing attribute $A$ using two queries $(Q_{1}, Q_{2})$, and using standard cracking \cite{idreos2007database}. Column $B$ is a copy of $A$ on which future queries are performed. $B$ is called the \emph{cracker column} in the literature. $T$ is called the \emph{cracker index}, and it tells future queries how $B$ is currently partitioned. These two structures together, $B$ and $T$, replace a traditional index. A node $\langle x,y\rangle$ in $T$ tells that \emph{all} elements of $B$ strictly larger than $x$ start at position $y$.}
	\label{figs:adaptive-index}
\end{figure}

In adaptive indexing, index creation is thought as a by-product of query execution; thus an index is built in a lazy manner as more queries are executed, see Figure~\ref{figs:adaptive-index}. The very first adaptive indexing algorithm, called \emph{standard cracking}, was presented in~\cite{idreos2007database}, and since then, adaptive indexing has received a considerable amount of research, see~\cite{kersten2005cracking,idreos2007updating,idreos2009self,graefe2011benchmarking,idreos2011merging,halim2012stochastic,graefe2012concurrency,schuhknecht2013uncracked,graefe2014transactional}. 

\subsection{Single-threaded adaptive indexing}\label{sec:intro:experiments}

In our recent paper~\cite{schuhknecht2013uncracked} we presented a thorough experimental study of all major single-threaded adaptive indexing algorithms. The experiments shown therein support the claim that (single-threaded) standard cracking~\cite{idreos2007database} still keeps being the algorithm any other new (single-threaded) algorithm has to improve upon, due to its simplicity and good accumulated runtime. For example, in the comparison of standard cracking~\cite{idreos2007database} with stochastic cracking~\cite{halim2012stochastic}, as seen from the experiments shown in~\cite{schuhknecht2013uncracked}, the latter is more robust, but the former is in general faster --- see Figure~4.c of~\cite{schuhknecht2013uncracked}. With respect to hybrid cracking algorithms~\cite{idreos2011merging}, the experiments of~\cite{schuhknecht2013uncracked} indicate that, although convergence towards full index improves in hybrid cracking algorithms, this improvement can be seen only after roughly 100--200 queries, before that, standard cracking performs clearly better --- see Figure~4.a of~\cite{schuhknecht2013uncracked}. Moreover, as also seen in Figure~8.a of~\cite{schuhknecht2013uncracked}, w.r.t.~total accumulated query time, standard cracking is faster up to around 8000 queries. In all cases, standard cracking is an algorithm that is much easier to implement and maintain.

In~\cite{schuhknecht2013uncracked} we also observed that pre-processing the input before standard cracking~\cite{idreos2007database} is used, significantly improves the convergence towards full index, robustness, and total execution time of standard cracking. This pre-processing step is just a range-partitioning over the attribute to be (adaptively) indexed. For simplicity we will refer from now on to this adaptive indexing technique simply as the \emph{coarse-granular index}, just as it is referred to in~\cite{schuhknecht2013uncracked}. From the experiments of~\cite{schuhknecht2013uncracked}, the improvement of the coarse-granular index over all other adaptive indexing techniques considered in~\cite{schuhknecht2013uncracked} can clearly be seen, see Figures 7.a and 7.b of~\cite{schuhknecht2013uncracked}. However, the coarse-granular index also incurs in a higher initialization time, which, as discussed previously, is in general not desirable.

\subsection{Representative single-threaded experiment}\label{sec:intro:experiment}

Taking the experiments of~\cite{schuhknecht2013uncracked} as a reference, we have run a small, but very representative, set of experiments considering what we believe are the three best adaptive indexing algorithms to date --- standard cracking~\cite{idreos2007database}, hybrid crack sort~\cite{idreos2011merging}, and the coarse-granular index~\cite{schuhknecht2013uncracked}. We additionally compare these methods with sorting algorithms. 

The purpose of these experiments is \emph{not} to reevaluate our own material of~\cite{schuhknecht2013uncracked}, but rather to make this paper from this point on self-contained. Also, these experiments represent the baselines for the multi-threaded algorithms shown later on --- these experiments are the entry point of all other experiments herein presented.

The source code used for this representative set of experiments is a tuned version of the code we used in~\cite{schuhknecht2013uncracked}. The workload under which we test the algorithms is defined in~\S~\ref{sec:intro:workload} below, and the experimental set-up is given in~\S~\ref{sec:experiments}. 

\begin{figure}[!htb]
	\begin{center}
		\includegraphics[scale=0.75]{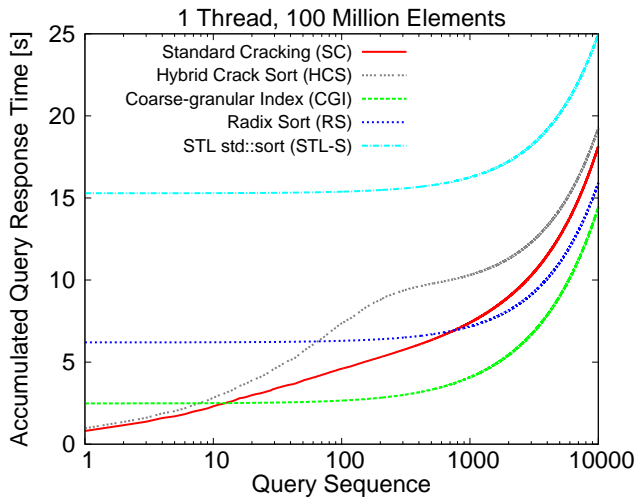}
	\end{center}
	\caption{Total time of the single-threaded algorithms over 10,000 random queries.}
	\label{figs:single-threaded}
\end{figure}

The results of this first set of experiments can be seen in Figure~\ref{figs:single-threaded}. The most important message, at this point, is that the reader can rest assured that these experimental results are fully consistent with the ones already found in the literature, in particular with our own work presented in~\cite{schuhknecht2013uncracked}. In particular, for a moderately large number of queries, we see that adaptive indexing algorithms achieve a high speedup over sorting algorithms, typically used for building full indexes. Thus, in single-threaded environments, adaptive indexing is a viable option in practice.

\subsection{Workload}\label{sec:intro:workload}

We assume a main-memory column-store with 100 million tuples (which we also increase to 1 billion tuples later on), filled with $4$-byte positive integers generated at random, w.r.t.~the perfect uniform distribution over the interval $\left[0, 2^{32}\right)$. Each entry in the cracker column is then represented as a pair $(key, rowID)$ of $4$-byte positive integers each. We are uniquely interested in analytical queries, thus we assume that no query performs updates on the database. To be more precise, we are only interested in the following type of queries: 

\begin{center}
	{\tt SELECT SUM($R.A$) FROM $R$ WHERE $q_{l}\leq R.A < q_{h}$}
\end{center}

That is, each query has to filter the data from $A$ and perform a simple aggregation over the set of results. The cracker column will always be clear from the context, thus, a query $q$ can then be represented simply as a pair $(q_{l}, q_{h})$. 
We perform $10,000$ queries generated uniformly at random (which we change to skewed distribution later on), each having its selectivity set at $1\%$. As in previous work, we assume that queries are independent from each other and are performed as they arrive. Also, we assume no idle time in the system, \emph{i.e.}, the cracker column is allocated immediately after the very first query arrives. Every measurement was run ten times. The input of the $i$-th run is the same for every algorithm, but any two different runs were generated completely independent from each other. The times reported for each algorithm are the average times over those ten runs. In all experiments we project only the index attribute. This means that no base table access is performed. In~\S~\ref{sec:intro:understanding} we elaborate more on the considered workload.

\subsection{The contributions of this paper}\label{sec:intro:contribution}

All algorithms tested in~\cite{schuhknecht2013uncracked} are single-threaded, and actually, almost \emph{all} adaptive indexing algorithms in the literature are designed for single-core systems. With multi-core systems not only on the rise, but actually well established by now, the idea of parallel adaptive indexing algorithms is very natural. 

To the best of our knowledge, the work presented in~\cite{graefe2012concurrency,graefe2014transactional} is the only (published) work, so far, that deals with adaptive indexing in multi-core systems. Therein, a parallel version of the standard cracking algorithm of~\cite{idreos2007database} is presented.

We are now ready to list the contributions of this paper:

\begin{enumerate}
	\item[(\num{1})] First and foremost, the most important contribution of this paper is the (critically) experimental evaluation of many parallel algorithms (7 to be precise), for adaptive indexing as well as for creating full indexes. This work can be considered the very first thorough experimental study of parallel adaptive indexing techniques.
	\item[(\num{2})] We describe three alternative parallel algorithms for adaptive indexing, one based on standard cracking~\cite{idreos2007database}, while the other two are based on the coarse-granular index presented in~\cite{schuhknecht2013uncracked}.
	\item[(\num{3})] We describe a hybrid parallel sorting algorithm. This algorithm greatly resembles the parallel radix sorting algorithms presented in~\cite{lee2002partitioned,maus2011parl,rashid2010analyzing}. Additionally, we present another method that further improves upon the hybrid parallel sorting algorithm by adding NUMA-awareness.
	\item[(\num{4})] We experimentally compare our algorithms with the algorithm of~\cite{graefe2012concurrency,graefe2014transactional}, as well as with the parallel version of radix sort presented in~\cite{maus2011parl}; which is therein reported to be very fast in practice.
	\item[(\num{5})] The initial set of experiments considered in this paper suggests that our algorithms significantly improve over the methods presented in~\cite{graefe2012concurrency,graefe2014transactional,maus2011parl}. Moreover, these experiments also seem to indicate that, as opposed to the story with single-threaded algorithms where sorting algorithms are no match in practice against adaptive indexing techniques, parallel sorting algorithms could become serious alternatives to parallel adaptive indexing techniques.
\end{enumerate}

Table~\ref{table:algorithms} gives an overview of all algorithms considered in this paper. Throughout the paper we mostly refer to the algorithms by their short names. Thus, for an easier distinction between single- and multi-threaded algorithms, the short name of \emph{each} multi-threaded algorithm starts \mbox{with a \algo{P}.}

\begin{table*}[!htb]
	\footnotesize
	\begin{center}
		\begin{tabular}{l c c l c}
			\hline
			\algorithm & {\tt Reference} & {\tt Description} & \moniker & {\tt Implementation used}\\
			\hline
			Standard cracking & \cite{idreos2007database} & \S~\ref{sec:algorithms:s-crack:single} &\SC{} & \cite{schuhknecht2013uncracked}\\
			Hybrid crack sort & \cite{idreos2011merging} & \cite{idreos2011merging} & \HCS{} & \cite{schuhknecht2013uncracked}\\
			Coarse granular index & \cite{schuhknecht2013uncracked} & \S~\ref{sec:algorithms:cg-idx:single} & \CGI{} & \cite{schuhknecht2013uncracked}\\
			Radix sort & \cite{maus2002arl} & \S~\ref{sec:algorithms:full:single} & \RS{} & \cite{schuhknecht2013uncracked}\\
			STL sort ({\tt C++})& \cite{stepanov1995standard} & \cite{stepanov1995standard} & \STL{} & {\tt C++ STL}\\
			\hline
			Parallel standard cracking & \cite{graefe2012concurrency,graefe2014transactional} & \S~\ref{sec:algorithms:s-crack:multi} & \PSC{} & \cite{graefe2012concurrency}\\
			Parallel radix sort & \cite{maus2011parl} & \S~\ref{sec:algorithms:full:multi} & \PRS{} & Ours\\			
			Parallel-chunked standard cracking & This paper & \S~\ref{sec:algorithms:s-crack:multi} & \PCSC{} & Ours\\
			Parallel coarse-granular index & This paper & \S~\ref{sec:algorithms:cg-idx:multi} & \PCGI{} & Ours\\
			Parallel-chunked coarse-granular index & This paper & \S~\ref{sec:algorithms:cg-idx:multi} & \PCCGI{} & Ours\\
			Parallel range-partitioned radix sort & This paper & \S~\ref{sec:algorithms:full:multi} & \PRPRS{} & Ours\\
			Parallel-chunked radix sort & This paper & \S~\ref{sec:algorithms:full:multi} & \PCRS{} & Ours\\
			\hline
		\end{tabular}
		\caption{All algorithms considered in this paper along their original references, their short names, and the implementations used in this paper.}
		\label{table:algorithms}
	\end{center}
\end{table*}

\subsection{How is this paper not to be understood}\label{sec:intro:understanding}

\begin{enumerate}
\item[(\num{1})] We are fully aware of \emph{all} prior work in adaptive indexing, including the very recent paper of Graefe \emph{et al.}~\cite{graefe2014transactional}. In this regard, it is \emph{not} the purpose of this paper to present parallel versions of \emph{every} single adaptive indexing algorithm in the literature. Our previous experimental study~\cite{schuhknecht2013uncracked} helped us to narrow down our options to the adaptive indexing techniques that show real potential for parallelization. Here we would also like to point out that in the beginning we also considered the hybrid crack sort algorithm of~\cite{idreos2011merging}, as it initially looked very promising. However, in Figure~\ref{figs:single-threaded} we can see that this algorithm starts catching up with standard cracking~\cite{idreos2007database} only after $10,000$ queries, but sorting catches up with standard cracking in less than $1,000$ queries. From that point on radix sort is actually faster than hybrid crack sort. From the observations of our single-threaded experiment, we decided on focusing our efforts in standard cracking~\cite{idreos2007database}, the coarse-granular index~\cite{schuhknecht2013uncracked}, and sorting algorithms.

\item[(\num{2})] We are also fully aware of the recent development in sorting algorithms (sort-merge) in the context of joins, see for example~\cite{kim2009sort,albutiu2012massively,balkesen2013multi} and references therein. In~\cite{balkesen2013multi}, well-engineered implementations of these sorting algorithms are used for the experiments therein presented. Unfortunately, the source code of these carefully-tuned implementations is not yet available at the time of this manuscript, and thus we could not test them against the sorting algorithms herein presented. Nevertheless, we would like to point out the following: It is \emph{not} the purpose of this paper to claim the fastest parallel sorting algorithm. It is our main purpose to compare, and put into context, \emph{good} parallel sorting algorithms with the best parallel adaptive indexing algorithms we could come up with. Faster parallel sorting algorithms would make the point even stronger that, in multi-core systems, parallel sorting algorithms could become serious alternatives to parallel adaptive indexing techniques.

\item[(\num{3})] Although the workload considered in this paper is fully equivalent to the ones already found in the literature, in particular to the one considered in~\cite{graefe2012concurrency,graefe2014transactional}, we are fully aware that it might not be realistic in practice. In a companion paper we will present results regarding a more realistic workload --- multi-selection-, multi-projections queries working on an entire table instead of only on a single index column. This in particular includes tuple-reconstruction. We consider this more realistic workload on a larger multi-core machine.
Nevertheless, we think that our simplified environment (experimental setup and workload) gives us a very good first impression of the effect of multi-threading in (adaptive) indexing.
\end{enumerate}

The implementations used in this paper are described in Table~\ref{table:algorithms} on an algorithm basis. These implementations will be freely available upon publication of the paper. 

\subsection{The structure of this paper}

The remainder of the paper is structured as follows: In~\S~\ref{sec:algorithms} we describe the algorithms to be tested. In~\S~\ref{sec:setup} we give our experimental setup. In~\S~\ref{sec:experiments} we give a precise definition of the workload used to test the algorithms, and show the results of our experiments. In~\S~\ref{sec:experiments:benchmarks:obs} we discuss the observed speedups. Later, in~\S~\ref{sec:additional} we also show the effect of varying different parameters of the workload, such as input size (\S~\ref{sec:experiments:input}), tuple configuration (\S~\ref{sec:experiments:config}), query access pattern (\S~\ref{sec:experiments:skewness}), distribution of input data (\S~\ref{sec:experiments:skewedInput}), and query selectivity (\S~\ref{sec:experiments:selectivity}). In~\S~\ref{sec:conclusions} we close the paper with some discussion and conclusions. 

Throughout the paper we have made our presentation as self-contained as possible.

\section{Algorithms}\label{sec:algorithms}

In this section we give a description of all parallel algorithms to be tested. For completeness, nevertheless, we also offer a description of the single-threaded algorithms used as a reference.

Throughout the paper we will denote by $A$ the original column (attribute) we want to perform queries on, by $B$ the corresponding cracker column, see Figure~\ref{figs:adaptive-index}, by $n$ the number of entries in $A$, and by $k$ the number of available threads. To make the explanation simpler, we will assume that $n$ is perfectly divisible by $k$. In the experiments, however, we do not make this assumption.

\subsection{Standard cracking}\label{sec:algorithms:s-crack}

\subsubsection{Single-threaded}\label{sec:algorithms:s-crack:single}

\algo{Standard cracking (\SC{}).} This is the first adaptive indexing algorithm that appeared~\cite{idreos2007database}, and also the most popular one due to its simplicity and good performance. This algorithm is by now well-known, and it can be described as follows: Let \mbox{$q = (q_{l}, q_{h})$} be a query, with $q_{l} < q_{h}$. In its simplest version, in order to filter out the results, standard cracking performs two partition steps of quicksort~\cite{hoare1962quicksort} over $B$, each one taking $q_{l}$ and $q_{h}$ as their pivots respectively. The split location of each partition step is then inserted into the cracker index, using $q_{l}$ and $q_{h}$ as the keys, so that future queries can profit from the work previous queries have already done, see Figure~\ref{figs:adaptive-index}. This incrementally improves query response times of standard cracking, as partitions become smaller over time.

The method of performing two partition steps is called $2$x-crack-in-two in~\cite{idreos2007database}, but the authors also observed that both partition steps can be combined into a single one, which they call crack-in-three. Experiments shown in~\cite{schuhknecht2013uncracked} suggest that $2$x-crack-in-two performs better than crack-in-three most of the time, and when it does not, the difference is small. Thus, the used implementation of standard cracking performs only $2$x-crack-in-two steps. Finally, it is important to know that the initialization cost of standard cracking is only that of creating the cracker column $B$, \emph{i.e.}, copying column $A$ onto $B$ before performing the very first query. However, it has been pointed out before in~\cite{idreos2007database} that, as to better take advantage of the creation of $B$, the very first partition (crack) can be materialized as the cracker column $B$ is created. In that way, after $B$ has been fully created, we can continue partitioning from the second crack on. Our implementation does precisely that. 

\subsubsection{Multi-threaded}\label{sec:algorithms:s-crack:multi}

\algo{Parallel standard cracking (\PSC{}).} In~\cite{graefe2012concurrency,graefe2014transactional}, a multi-thread\-ed version of standard cracking was shown. To describe this multi-threaded version it suffices to observe that, in standard cracking, as more queries arrive, they potentially partition independent parts of $B$, and thus, they can be performed in parallel. 

When a query comes, it has to acquire two write locks on the border partitions, while \emph{all} partitions in between are protected using read locks. When two or more queries have to partition, or aggregate over the same part of $B$, read and write locks are used over the relevant parts. That is, whenever two or more queries $q_{1},\ldots, q_{r}$, $r\geq 2$, want to partition the same part of $B$, a write lock is used to protect that part; say $q_{i}$, $1\leq i\leq r$, obtains the lock and partitions while the other queries wait for it to finish. After $q_{i}$ has finished, the next query $q_{j}$, $i\neq j$, acquiring the lock has to reevaluate what part it will exactly crack, as $q_{i}$ has modified the part all queries $q_{1},\ldots, q_{r}$ were originally interested in. Clearly, as more queries are performed, the number of partitions in $B$ increases, and thus also the probability that more queries can be performed in parallel. This is where the speedup of this multi-threaded version over the single-threaded version stems from. If two or more queries want to aggregate over the same part of $B$, then they all can be performed in parallel, as they are not physically reorganizing any data. However, if one query wants to aggregate over a part of $B$ that is currently being partitioned by another query, then the former has to wait until the latter finishes, as otherwise the result of the aggregation might be incorrect. Also, all queries work with the same cracker index, thus a write lock occurs each time the cracker index is updated. 

As the initialization time of \PSC{} we consider the time it takes to copy $A$ onto $B$ in parallel. That is, for $k$ available threads, we divide $A$ and $B$ into $k$ parts, and assign exactly one part to each thread. Every thread then copies its corresponding part from $A$ to $B$. 

We immediately see two drawbacks with this multi-threaded version of standard cracking, which will also become apparent in the experiments: (\num{1}) The effect of having multiple threads will be visible only after the very first executed query has partitioned $B$. Before that, $B$ consists of only one partition, and all other queries will have to wait for this very first query to finish. That is, the very first crack locks the whole column. (\num{2}) Locking incurs in unwanted time overheads.

To address these concerns we present in this paper another version of parallel standard cracking, which as we will see, seems to perform quite good in practice, in particular, better than \PSC{}.

\algo{Parallel-chunked standard cracking (\PCSC{}).} After copying $A$ onto $B$ in parallel, as in \PSC{}, we (symbolically) divide $B$ into $k$ parts, each having $\frac{n}{k}$ elements, and every thread will be responsible for exactly one of these parts. Now, \emph{every} query will be executed by \emph{every} thread on its corresponding part, and \emph{every} thread will aggregate its results to a local variable assigned to it. At the end a single thread aggregates over all these local variables. 

It is crucial for the performance of \PCSC{} to ensure complete independence between the individual parts. Any data that is unnecessarily shared among them can lead to \emph{false sharing effects} (propagation of cache line update to a core although the update did not affect its part of the shared cache line) and \emph{remote accesses} to memory attached to another socket. Thus, each part maintains its own structure of objects, containing its local data, cracker index, histograms, and result aggregation variables. Furthermore, by aligning all objects to cache lines, we ensure to avoid any shared resources, and each thread can process its part in complete independence from the remaining ones. Finally, we also pin threads (during its lifetime) to physical cores. This gives the strong hint that, when a thread instantiates its part --- and all variables around it, it should do so in its NUMA region.

\subsection{Coarse-granular index}\label{sec:algorithms:cg-idx}

\subsubsection{Single-threaded}\label{sec:algorithms:cg-idx:single}

\algo{Coarse-granular index (\CGI{}).} In~\cite{schuhknecht2013uncracked}, a new adaptive indexing technique, therein called the \emph{coarse-granular index}, was presented. This technique range-partitions $B$ as a pre-proc\-ess\-ing step\footnote{If there is a bias in the distribution of the keys, an equi-depth partition could be used instead of a range partition; at the expense of more pre-processing time.}. That is, the range of values of the keys of $B$ is divided into \mbox{$r\geq 2$} buckets\footnote{In the core part of our experiments, as in the ones shown in~\cite{schuhknecht2013uncracked}, we set \mbox{$r = 1024$}, since that was the value for which the  best performance was observed. When increasing the size of the input by a factor of ten we set \mbox{$r = 8192$}.}, such that the first bucket contains the first $\frac{n}{r}$ largest values in $B$, the second bucket contains the second $\frac{n}{r}$ largest values, and so on. Again, for the sake of explanation, we assume that $n$ is divisible by $r$. In the experiments, however, we do not make this assumption. The keys inside each bucket are in any arbitrary order. Once $B$ has been range-partitioned, the position where each bucket ends is inserted in a cracker index $T$, and standard cracking \SC{} is run on $B$ to answer the queries; taking $T$ into consideration. 

Range partitioning $B$ gives standard cracking a huge speed-up over standard cracking alone, see Figure~\ref{figs:single-threaded}. As it turns out, this range partitioning can be done very fast; the elements of $A$ can be range-partitioned while being copied to $B$. This requires only two passes over $A$. Also, as we pointed out in~\cite{schuhknecht2013uncracked}, \CGI{} converges faster towards full index than \SC{}, and is also more robust w.r.t. skewed queries. However, \CGI{} incurs in relatively high initialization time w.r.t.~\SC{}, see Figure~\ref{figs:single-threaded}. There, it can be seen that \SC{} can perform around $10$ queries while \CGI{} is still range-partitioning, after that, \CGI{} pays off already.

\subsubsection{Multi-threaded}\label{sec:algorithms:cg-idx:multi}

In this paper we present two parallel versions of \CGI{}. For the first one it suffices to observe that \CGI{} is nothing but a range partitioning as a pre-processing step to \SC{}.  Thus, for the first parallel version of \CGI{}, we show how to do a range partitioning in parallel. Afterwards we simply run \PSC{} to answer the queries, taking into consideration that $B$ is now range-partitioned. Our method to build a range partition in parallel requires no synchronization among threads, which of course helps to improve its performance. 

\begin{figure}[!htb]
	\begin{center}
		\includegraphics[scale=0.45]{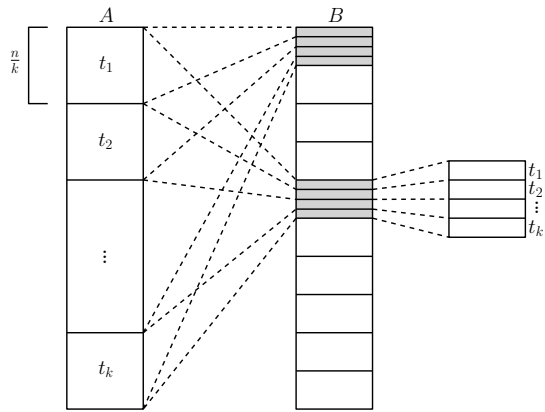}
	\end{center}
	\caption{In parallel coarse-granular index (\PCGI{}), column $A$ is (symbolically) divided into $k$ parts of $\frac{n}{k}$ elements each. Cracker column $B$ is range-partitioned using $r$ buckets. Every bucket of the range partition of $B$ is also (symbolically) divided into $k$ parts. Thread $t_{i}$, $1\leq i\leq k$, writes to the $i$-th part of every bucket.}
	\label{figs:p-cgi}
\end{figure}

\algo{Parallel coarse-granular index (\PCGI{}).} The main idea behind the construction is very simple. Column $A$ is (symbolically) divided into $k$ parts, of $\frac{n}{k}$ elements each. Thread $t_{i}$, $1\leq i\leq k$, gets assigned the $i$-th part of $A$ and it writes its elements to their corresponding buckets in the range partition on $B$, using $r\geq 2$ buckets. In order to do so, and not to incur in any synchronization overhead, \emph{every} bucket of the target range partition on $B$ is (symbolically) divided into $k$ parts as well, so that thread $t_{i}$ writes its elements in the $i$-th part of \emph{every} bucket. Thus, clearly, any two threads read their elements from independent parts of $A$ and write also to independent parts on $B$, see Figure~\ref{figs:p-cgi}. All this can be implemented in a way that all but one step are done in parallel\footnote{This step is the aggregation of an histogram used by all threads.}, and every thread gets roughly the same amount of work. This, as we will see, helps to improve performance as the number of threads increases. 

The second parallel version of \CGI{} does not range-partition $B$. Instead, it works in the same spirit as \PCSC{}, thus its name.

\algo{Parallel-chunked coarse-granular index (\PCCGI{}).} Symbolically divide $A$ again into $k$ parts, of $\frac{n}{k}$ elements each, and assign the \mbox{$i$-th} part to the $i$-th thread. Each thread $t_{i}$, $1\leq i\leq k$, range partitions its part using \CGI{}, materializing it onto $B$. Thus, $B$ is also (symbolically) divided into $k$ parts. Having done this chunked range partition of $B$, thread $t_{i}$ keeps being responsible for the $i$-th range partition of $B$. When a query arrives, each thread executes \SC{} on its part and aggregates its result in a global variable, for which we again use a write lock to avoid any conflicts that might occur during aggregation. Again, as for \PCSC{}, we ensure that no resources are shared among the parts. All objects are cache-line-aligned and no trips to the remote memory are necessary at any place. The initialization time for each one of these algorithms, \CGI{}, \PCGI{}, and \PCCGI{}, is the time it takes to materialize the necessary range partitions on $B$.

\subsection{Full indexing}\label{sec:algorithms:full}

Up to now we have only described algorithms for adaptive indexing. However, in order to see how effective those algorithms really are, we have to compare them against full indexes. In~\cite{schuhknecht2013uncracked} it was observed that, when the selectivity of range queries is not extremely high, as in our case, more sophisticated indexing data structures such as AVL-trees, B$^{+}$-trees, ART~\cite{leis2013adaptive}, among others, have no significant benefit over \mbox{full sort + binary search + scan} for answering queries, as the scan cost (aggregation) of the result dominates the overall query time. Therefore, for our study, we regard sorting algorithms as a direct equivalent of full indexing algorithms. 

\subsubsection{Single-threaded}\label{sec:algorithms:full:single}

\algo{Radix sort (\RS{}).} The single-threaded sorting algorithm used in our experiments is the one presented in~\cite{maus2002arl}. This algorithm was also used in~\cite{schuhknecht2013uncracked}, where it was reported to be very fast. This algorithm is a recursive {\textbf{M}}ost {\textbf{S}}ignificant {\textbf{D}}igit radix sort, called left radix sort in~\cite{maus2002arl}. This radix sort differs from the traditional {\textbf{L}}east {\textbf{S}}ignificant {\textbf{D}}igit radix sort~\cite{donald1999art} not only in that the former is MSD and the latter LSD, but also in that the former is in place and not stable, while the latter requires an extra array of size $n$, but is stable. Experiments shown in~\cite{maus2002arl} indicate that this MSD radix sort is faster than the traditional LSD radix sort when the input is large.

What the algorithm of~\cite{maus2002arl} does to work in place is the following: The input gets (symbolically) divided into $r = 2^{m}$ buckets, where $m$ is the number of bits of the sorting digits. Thus, $r$ represents the number of different values of the sorting digits. Now, instead of exchanging the keys between two arrays according to the value of the keys in the sorting digit, as in a traditional radix sort, the algorithm of~\cite{maus2002arl} works in permutation cycles \`{a} la Cuckoo~\cite{Pagh:2001aa}. That is, it places an element in its correct bucket, w.r.t.~the value of its sorting digit. In doing so, it evicts another element which is then placed in its correct bucket, as so on and so forth. Eventually, an element gets placed in the bucket of the very first element that initiated this permutation cycle. Then, the next element that has not been moved yet starts another permutation cycle, and so on. Eventually, all elements get moved to their corresponding buckets. At that point, the algorithm recurs in \emph{each} of the $r$ bucket that the input was (symbolically) divided into. This ensures that all the work done previously is not destroyed. When the number of elements in a bucket is small enough, the algorithm stops the recursion and uses insertion sort instead. This step improves performance in practice. In our implementation we treat the numbers as four-digit numbers --- radix-$2^{8}$ numbers. That is, $r = 256$. Thus, only four passes are necessary for sorting. Figure~\ref{figs:single-threaded} presents how \RS{} performs against \SC{}, \CGI{}, and \STL{}. This last algorithm is the STL sort of the {\tt C++} standard library~\cite{stepanov1995standard}, which we just use as an additional reference.

The take-home message of the experiments shown in Figure~\ref{figs:single-threaded} is that single-threaded adaptive indexing algorithms greatly outperform good sorting algorithms, for a moderately large number of queries. Thus, adaptive indexing is a viable option in practice.

\subsubsection{Multi-threaded}\label{sec:algorithms:full:multi}

The same author of~\cite{maus2002arl} presented in~\cite{maus2011parl} a parallel MSD radix sort, which is very similar to the algorithm presented in~\cite{lee2002partitioned}. We noticed in our experiments that this parallel MSD radix sort is very fast. Therefore, we decided to include it in this study. Its description is the following.

\begin{figure}[!htb]
	\begin{center}
		\includegraphics[scale=0.45]{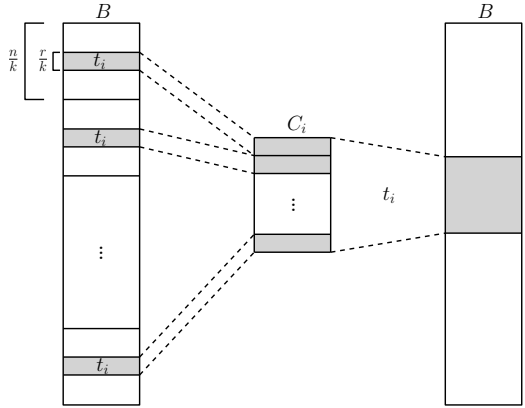}
	\end{center}
	\caption{To the left, every part of $B$, of $\frac{n}{k}$ elements each, is already sorted w.r.t.~the most significant digit. Thread $t_{i}$, $1\leq i\leq k$, now copies the $i$-th part, of each of the $k$ parts that $B$ is (symbolically) divided into, to a temporary array $C_{i}$. This $i$-th part consists of $\frac{r}{k}$ buckets created by \RS{} in the very first step. Thread $t_{i}$ now sorts $C_{i}$ w.r.t.~to the remaining sorting digits. After being sorted, $C_{i}$ gets copied to the $i$-th part of $B$ by $t_{i}$, shown to the right. After all threads have finished, $B$ is now fully sorted.}
	\label{figs:p-rs}
\end{figure}

\algo{Parallel radix sort (\PRS{}).} Let $m$ and $r$ be as for \RS{}. The main idea of the algorithm is the following: Column $A$ gets first copied onto $B$. Then $B$ gets (symbolically) divided into $k$ parts, each consisting of $\frac{n}{k}$ elements. Thread $t_{i}$, $1\leq i\leq k$, gets assigned the $i$-th part of $B$, and it sorts its elements using \RS{} \emph{only} w.r.t.~the most significant digit. That is, using only the $8$ most significant bits of each number. After all threads have finished, thread $t_{i}$, \mbox{$1\leq i\leq k$}, will be responsible for the set of numbers contained in buckets $\left[(i - 1)\cdot\frac{r}{k} + 1, i\cdot\frac{r}{k}\right]$ of \emph{each} of the $\frac{n}{k}$ parts that $B$ is divided into, see Figure~\ref{figs:p-rs}. These buckets are created by \RS{} when it was first run on each of the $\frac{n}{k}$ parts that $B$ is divided into. Thus, $t_{i}$ is now actually responsible for the globally $i$-th largest values in $B$. Thread $t_{i}$ now fetches all these numbers from all the parts that $B$ is divided into, and copies them onto a temporary array $C_{i}$, where it then sorts these numbers using \RS{} w.r.t.~the remaining sorting digits, \emph{i.e.}, starting from the second most significant digit, see Figure~\ref{figs:p-rs}. Having sorted $C_{i}$, thread $t_{i}$ now copies $C_{i}$ back to $B$, taking into consideration that $C_{i}$ holds the $i$-th largest values in $B$. The correct offset of where to copy $C_{i}$ in $B$ can easily be computed with an histogram, which is the only sequential computation in the algorithm, the rest is performed in parallel. Also, every thread gets roughly the same amount of work. This helps to improve the performance as the number of threads increases.

As we will see, \PRS{} is very fast in practice. However, it requires extra space for the temporary arrays $C_{i}$, $1\leq i\leq k$, which is definitely a drawback, since space becomes critical as the size of the input increases. Therefore, due to: (\num{1}) The aforementioned drawback of \mbox{\PRS{}}. (\num{2}) The good performance of building a range partition in parallel, and (\num{3}) The good performance of \RS{}. The following hybrid algorithm suggests itself:

\algo{Parallel range-partitioned radix sort (\PRPRS{}).} Build a range partitioning in parallel on $B$, as in \PCGI{}. If the number of buckets in the range partitioning is $r = 2^{m}$, then it is not hard to see that the elements of $B$ are now sorted w.r.t.~the $m$ most significant bits. Now, split the buckets of the range partitioning evenly among all $k$ threads. Each thread then sorts the elements assigned to it on a bucket basis using \RS{}, but starting from the \mbox{$(m+1)$-th} most significant bit; remember that \RS{} is a MSD radix sort. Since $B$ is range-partitioned, and sorted w.r.t.~the $m$ most significant bits, calling \RS{} on each bucket clearly fully sorts $B$ in-place. 

As we will see, \PRPRS{} performs already better than \PRS{}. However, both sorting algorithms suffer from a large amount of NUMA effects. To alleviate this we present the following algorithm:

\algo{Parallel-chunked radix sort (\PCRS{}).} Symbolically divide $A$ into $k$ chunks, of $\frac{n}{k}$ elements each, and assign the $i$-th part to the $i$-th thread. Each thread $t_i$, $1 \leq i \leq k$, range partitions its part using \CGI{}, materializing it onto the corresponding chunk of $B$. Afterwards, each thread $t_i$ reuses the histogram of its chunk to sort these partitions using \RS{} starting from the $(m+1)$-th most significant bit (the range-partitioning already sorts w.r.t.~the $m$ most significant bits). Thus, \PCRS{} basically applies the concept of \PRPRS{} to $k$ chunks. As for all other chunked methods, we ensure that the chunks do not share any data structures and that all objects are again cache-line-aligned. Thus, the threads work completely independent from each other.

The initialization time for the parallel sorting algorithms \mbox{\PRS{}}, \mbox{\PRPRS{}}, and \mbox{\PCRS{}} is clearly the time it takes them to sort. After that, for \PRS{} and \PRPRS{}, the queries can be answered in parallel using binary search; every thread will answer a different query (inter-query parallelism). In contrast to that, the chunked \PCRS{} answers the individual queries in parallel (intra-query parallelism) by querying the chunks concurrently. 

Table~\ref{table:algorithms} in~\S~\ref{sec:intro:contribution} constitutes a summary of the algorithms considered in this paper. We have decided to leave linear scan and hybrid cracking algorithms out of the main presentation of our experiments due to their high execution time, even in parallel.

\section{Experimental setup}\label{sec:setup}

Our test system consists of a single machine having two Intel Xeon E5-2407 running at 2.20 GHz. Each processor has four cores, and thus the machine has eight (hardware) threads. Hyper-threading and turbo-boost is not supported by the processors. The L1 and L2 cache sizes are 32 KB and 256 KB respectively per core. The L3 cache is shared by the four cores in the same socket and has size 10 MB. The machine has a total of 48 GB of shared RAM. The operating system is a 64-bit version of Linux. All programs are implemented in {\tt C++} and compiled with the Intel compiler {\tt icpc 14.0.1}~\cite{intel-compiler} with optimization {\tt -O3}.

\section{Core experiments}\label{sec:experiments}

We start by showing the core part of our experiments under the workload defined in~\S~\ref{sec:intro:workload}. To better observe the effect of the number of threads in all algorithms, we ran the experiments with $2, 4$, and $8$ threads.

\subsection{Benchmarks considered}\label{sec:experiments:benchmarks}

The experiments for one thread correspond to our small representative experiment discussed in~\S~\ref{sec:intro:experiment}, and whose results are shown in Figure~\ref{figs:single-threaded}. The two benchmarks considered in this evaluation are: Initial response time, and total execution time.

From the initial response time we get an idea on how long it takes the algorithms to start answering queries once they have been asked to. That is, we consider the initialization time, as explained in~\S~\ref{sec:algorithms} for each algorithm, plus the time taken by the very first query. For single-threaded algorithms, initial response time is the strongest point in favor of standard cracking, and against sorting algorithms and other adaptive indexing techniques. From Figure~\ref{figs:single-threaded} we can observe that the single-threaded version of standard cracking, \SC{}, can perform around $1,000$ queries while single-threaded version of radix sort, \RS{}, catches up. After that threshold \RS{} becomes faster than \SC{}. The comparison against STL sort is even worse. The story looks very different when comparing the single-threaded version of the coarse-granular index, \CGI{}, against \SC{}. From the same Figure~\ref{figs:single-threaded} we observe that \SC{} can perform only about $10$ queries while \CGI{} is building its range partitioning. After that threshold the coarse-granular index is already faster.

\begin{figure}[!htb]
	\centering
	\includegraphics[scale=0.75]{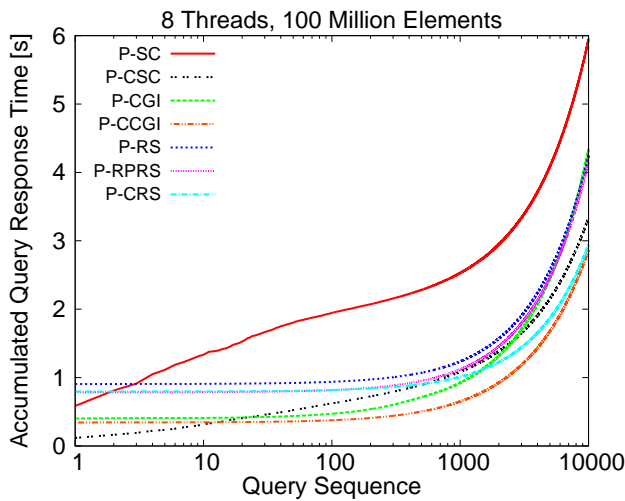}
	\caption{Total time of the 8-threaded algorithms over 10,000 random queries.}
	\label{figs:8-threaded}
\end{figure}

From the total execution time we clearly obtain the overall speedup of the multi-threaded algorithms over their single-threaded counterparts. We show here only the figure corresponding to the $8$-threaded experiment, see Figure~\ref{figs:8-threaded}. 

\subsection{Single-threaded vs. multi-threaded}

Based on the results we obtained in the single-threaded experiment, see Figure~\ref{figs:single-threaded}, we can now analyze the behavior of the methods using multiple threads.  As a first comparison we show how the parallel algorithms perform w.r.t.~their single-threaded counterparts. In this section we only present the results of the experiments, the speedups observed are discussed in details in~\S~\ref{sec:experiments:benchmarks:obs}.

In Table~\ref{table:sc-comp} we compare all standard cracking algorithms described in~\S~\ref{sec:algorithms:s-crack}. In Table~\ref{table:cgi-comp} we compare all coarse-granular algorithms described in~\S~\ref{sec:algorithms:cg-idx}. In Table~\ref{table:sort-comp} we compare all sorting algorithms described in~\S~\ref{sec:algorithms:full}.  In all tables, times are reported in seconds. The speedup reported is w.r.t.~the single-threaded version of the respective algorithm. The absolute best times among all algorithms, w.r.t.~the number of threads and benchmarks, are highlighted in blue. 

\begin{table}[!htb]
	\footnotesize
	\begin{center}
		\begin{tabular}{l c c c c c}
			\hline
			\algorithm & \threads & $\benchb$ & \speedup & $\bencha$ & \speedup\\
			& & {\tt time} & & {\tt time} &\\
			\hline
			\hline
			\SC{} & 1 & $0.8076$ & $1\times$ & $18.14$ & $1\times$\\
			\hline
			\PSC{} & 2 & $0.6617$ & $1.22\times$ & $13.82$ & $1.313\times$\\
			\PCSC{} & 2 & \cellcolor{color}$0.4067$ & $1.986\times$ & $9.051$ & $2.004\times$\\
			\hline
			\PSC{} & 4 & $0.5744$ & $1.406\times$ & $8.225$ & $2.205\times$\\
			\PCSC{} & 4 & \cellcolor{color}$0.2093$ & $3.859\times$ & $4.95$ & $3.664\times$\\
			\hline
			\PSC{} & 8 & $0.5863$ & $1.377\times$ & $5.957$ & $3.044\times$\\
			\PCSC{} & 8 & \cellcolor{color}$0.1178$ & $6.859\times$ & $3.344$ & $5.423\times$\\
			\hline
		\end{tabular}
		\caption{Comparison of standard cracking algorithms. Times are shown in seconds.}
		\label{table:sc-comp}
	\end{center}
\end{table}

\begin{table}[!htb]
	\footnotesize
	\begin{center}
		\begin{tabular}{l c c c c c}
			\hline
			\algorithm & \threads & $\benchb$ & \speedup & $\bencha$ & \speedup\\
			& & {\tt time} & & {\tt time} &\\
			\hline
			\hline
			\CGI{} & 1 & $2.483$ & $1\times$ & $14.43$ & $1\times$\\
			\hline
			\PCGI{} & 2 & $1.488$ & $1.668\times$ & $10.9$ & $1.324\times$\\
			\PCCGI{} & 2 & $1.243$ & $1.997\times$ & \cellcolor{color}$7.212$ & $2.001\times$\\
			\hline
			\PCGI{} & 4 & $0.7456$ & $3.33\times$ & $6.086$ & $2.371\times$\\
			\PCCGI{} & 4 & $0.6293$ & $3.946\times$ & \cellcolor{color}$4$ & $3.608\times$\\
			\hline
			\PCGI{} & 8 & $0.4032$ & $6.158\times$ & $4.345$ & $3.321\times$\\
			\PCCGI{} & 8 & $0.3436$ & $7.226\times$ & \cellcolor{color}$2.867$ & $5.033\times$\\
			\hline
		\end{tabular}
		\caption{Comparison of coarse-granular index algorithms. Times are shown in seconds.}
		\label{table:cgi-comp}
	\end{center}
\end{table}

\begin{table}[!htb]
	\footnotesize
	\begin{center}
		\begin{tabular}{l c c c c c}
			\hline
			\algorithm & \threads & $\benchb$ & \speedup & $\bencha$ & \speedup\\
			& & {\tt time} & & {\tt time} &\\
			\hline
			\hline
			\RS{} & 1 & $6.201$ & $1\times$ & $15.91$ & $1\times$\\
			\hline
			\PRS{} & 2 & $3.219$ & $1.926\times$ & $9.118$ & $1.745\times$\\
			\PRPRS{} & 2 & $2.894$ & $2.143\times$ & $9.833$ & $1.618\times$\\
			\PCRS{} & 2 & $2.792$ & $2.221\times$ & $8.038$ & $1.98\times$\\
			\hline
			\PRS{} & 4 & $1.611$ & $3.85\times$ & $5.381$ & $2.957\times$\\
			\PRPRS{} & 4 & $1.467$ & $4.227\times$ & $5.306$ & $2.999\times$\\
			\PCRS{} & 4 & $1.485$ & $4.176\times$ & $4.453$ & $3.574\times$\\
			\hline
			\PRS{} & 8 & $0.9063$ & $6.842\times$ & $4.245$ & $3.748\times$\\
			\PRPRS{} & 8 & $0.7846$ & $7.903\times$ & $4.163$ & $3.823\times$\\
			\PCRS{} & 8 & $0.797$ & $7.78\times$ & $2.94$ & $5.413\times$\\
			\hline
		\end{tabular}
		\caption{Comparison of sorting algorithms. Times are shown in seconds.}
		\label{table:sort-comp}
	\end{center}
\end{table}

It comes as no surprise that multi-threaded versions of the algorithms performed better than the single-threaded ones. With respect to absolute initial response times, we can see that \mbox{\PCSC{}} is the fastest algorithm, followed by \mbox{\PSC{}} and \mbox{\PCCGI{}}. Yet, it is quite interesting to see that sorting algorithms profited from parallelism the most --- the highest speedups are observed there, up to $7.903\times$ speedup for eight threads. That is, we are now able to build a full index over the main column $A$ eight times faster. From these speedups we can observe that \PRPRS{} is highly CPU-bound --- it shows linear speedups w.r.t.~the number of threads (more on this in~\S~\ref{sec:experiments:benchmarks:obs}). Thus, even higher speedups are expected on systems with an even larger number of (logical) threads. 

With respect to total execution time, the best algorithm turned out to be \mbox{\PCCGI{}}, followed rather close by \mbox{\PCRS{}} and \mbox{\PCSC{}} --- the chunked methods turned out to be the best w.r.t.~total execution time, where we can see a five-fold increase in speed for eight threads.

\section{Discussing the speedups}\label{sec:experiments:benchmarks:obs}

\begin{figure*}[!htb]
	\centering
	\subfloat[The highest rate at which memory is consumed is 36.773~GB/s. The colors indicate the different stages of the algorithm (from left to right): (\num{1}) Histogram creation, (\num{2}) Materializing the range-partitioning, (\num{3}) Radix sorting, and (\num{4}) Query answering.]{\label{figs:pcrs8}\includegraphics[width=\columnwidth]{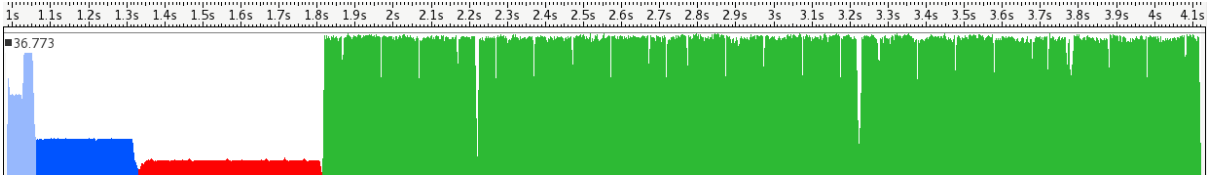}}
	\qquad
	\subfloat[The highest rate at which memory is consumed is 31.111~GB/s. The colors indicate the different stages of the algorithm (from left to right): (\num{1}) Copying the input to the cracker column, (\num{2}) Cracking and query answering. In the very beginning, up to around the \mbox{$3$-seconds} mark, waiting times hinder scalability. After that threshold there are enough partitions in the cracker column so that all threads can work mostly concurrently.]{\label{figs:psc8}\includegraphics[width=\columnwidth]{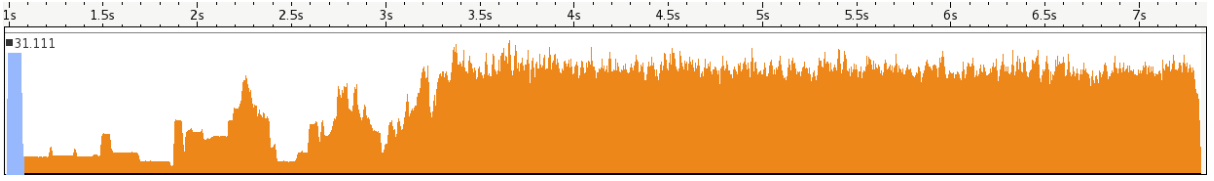}}
	\qquad
	\subfloat[The highest rate at which memory is consumed is 40.196~GB/s, which is actually the limit bandwidth of our system. The colors indicate the different stages of the algorithm (from left to right): (\num{1}) Histogram creation, (\num{2}) Sorting w.r.t.~the eight most significant bits, (\num{3}) Copying to final space (to contain the final sorted sequence), (\num{4}) Sorting in the final space w.r.t.~the remaining bits, and (\num{5}) Query answering.]{\label{figs:prs8}\includegraphics[width=\columnwidth]{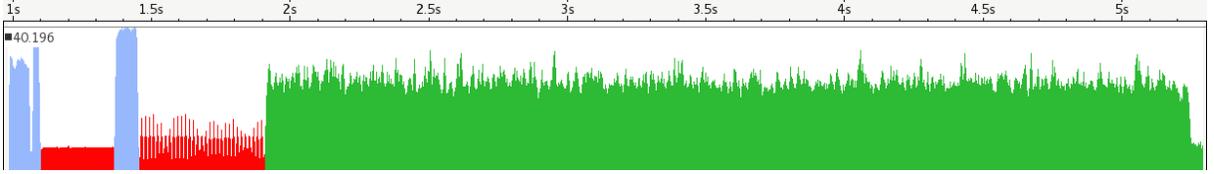}}
	\qquad
	\subfloat[The highest rate at which memory is consumed is 32.213~GB/s. The colors indicate the different stages of the algorithm (from left to right): (\num{1}) Histogram creation, (\num{2}) Materializing the range-partitioning, (\num{3}) Radix sorting, and (\num{4}) Query answering.]{\label{figs:prprs8}\includegraphics[width=\columnwidth]{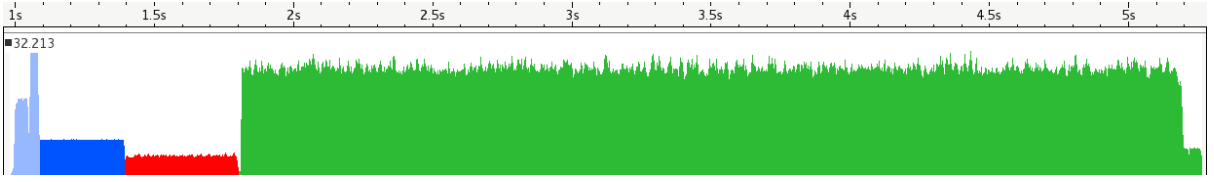}}
	\caption{Memory bandwidth utilization of the eight-threaded versions of (a) \mbox{\PCRS{}}, (b) \mbox{\PSC{}}, (c) \mbox{\PRS{}}, and (d) \mbox{\PRPRS{}}.}
	\label{figs:8-threadedbw}
\end{figure*}

As we explained in~\S~\ref{sec:algorithms}, most of our algorithms are lock-free, and are highly parallelizable. Still, we can observe that for $k$ threads, the speedups obtained are not always $k$-folded. To understand this phenomenon, we have profiled our algorithms with Intel Vtune Amplifier XE 2013~\cite{intel-vtune}. This tool allows us to gather all sorts of information about a running program, such as consumed memory bandwidth (GB/s), cpu utilization, cache misses, lock contention, among many others. We have carefully analyzed \emph{all} the considered algorithms for \emph{all} considered number of threads. Here, however, we do not discuss \emph{each} single algorithm in each one of the considered configurations, we will rather pick a representative subset of all configurations and give the explanations for them. Finally, we would like to point out that profiling runs of the algorithms are slower than the regular runs, as profiling code is added for this purpose.

\subsection{Chunked algorithms}

All chunked algorithms, \emph{i.e.}, \PCSC{}, \PCCGI{}, and \mbox{\PCRS{}}, share a large portion of code, and their implementations, as we already pointed out, are NUMA-aware. That is, we give the system the strong hint that each thread should work on its chunk in its own NUMA region by pinning the working threads to physical cores of the system. For example, for four threads, the first two threads (1, 2) will be pinned in one socket and the other two (3, 4) will be pinned in the other socket. For eight threads, the first four threads (1, 2, 3, 4) will be pinned in a socket, and the other four (5, 6, 7, 8) will be pinned in the other socket. We also avoid false-sharing among threads by aligning the working data of a thread with the boundary of the cache lines, so that for any two threads, the working data is found on different cache lines, and thus one thread does not invalidate the cache line of any other thread.

Now, let us consider the eight-threaded version of \mbox{\PCRS{}}, see Table~\ref{table:sort-comp}, and let us focus on total running time. There, we see that \mbox{\PCRS{}} achieves a speedup of only $5.413\times$, when ideally it should achieve a speedup of $8\times$. How this algorithm utilizes the memory bandwidth of the system can be seen in Figure~\ref{figs:pcrs8}, as reported by Vtune. The highest rate at which memory was consumed by \mbox{\PCRS{}} is $36.773$ GB/s. We measured the combined memory bandwidth of our system to be around $40$~GB/s, that is, $20$~GB/s per socket. So we can conclude that memory bandwidth is not the bottleneck of the algorithm. The real bottleneck is the following. In order to obtain $8\times$ speedup, we should obtain $4\times$ speedup from each socket (four threads per socket). The algorithm is highly parallelizable. However, we measured the speedup of each socket to be only $2.9\times$. We measured this by scheduling a four-threaded version of \mbox{\PCRS{}} in one socket only. The discrepancy between the expected $4\times$ speedup and the obtained $2.9\times$ speedup comes from the fact that the four threads are now sharing one L3 cache, the one corresponding to the socket. If this L3 cache was four times bigger, the configuration would be equivalent to the ideal configuration of having four sockets, each one having its own large L3 cache, and thus the processors would benefit from the larger cache sizes. We simulate this experiment with a two-threaded version of \mbox{\PCRS{}}, as we do not have a four-socket system. When this version is scheduled on one socket only, the number of cache misses\footnote{The counters quantified by Vtune are \mbox{\scriptsize{\tt OFFCORE\_RESPONSE.DEMAND\_DATA\_RD.LLC\_MISS.LOCAL\_DRAM}}\ , \mbox{\scriptsize{\tt OFFCORE\_RESPONSE.DEMAND\_DATA\_RD.LLC\_MISS.REMOTE\_DRAM}}} is on average (over ten runs of the algorithm) $231,060,000$ for this socket, as reported by Vtune. When this version is scheduled on two different sockets, the number of cache misses per socket is on average $102,500,000$. Overall there are $26,060,000~(\approx 12\%)$ more cache misses on the former configuration than in the latter, and this of course affects the scalability, going from $2\times$ speedup in the latter configuration, to $1.8\times$ speedup in the former. 

This explains the $2.9\times$ speedup of a single socket executing four threads. This also means that the achievable speedup by fully using both sockets (4 threads per socket) is about $2\cdot 2.9\times = 5.8\times$. Yet, we are obtaining $5.4\times$, as reported in Table~\ref{table:sort-comp}. This last discrepancy is a NUMA effect. Even though our implementations are fully NUMA-aware, at certain times during the execution of the algorithm, the system allocates temporary variables in different NUMA regions --- even though all threads are pinned to physical cores, and they should instantiate their data in their NUMA region. That is, every once in a while there are local cache misses that are served by data from a different NUMA region --- we noticed that roughly $1\%$ of the total number of caches misses come from a different NUMA region, this all can be seen in Vtune. It is well-known that bringing data from a different NUMA region is more expensive than accessing the same NUMA region. Thus this slightly slows down the algorithm.

We performed the very same analysis on \mbox{\PCSC{}} and \mbox{\PCCGI{}} and we noticed the same effect --- which was to be expected since all those methods share a large portion of code. The overall numbers are different nevertheless, and for brevity we will not show them here. 

With these arguments we are able to explain the speedups of \emph{all} chunked methods: \mbox{\PCSC{}}, \mbox{\PCCGI{}}, and \mbox{\PCRS{}}. We will now proceed to analyze \mbox{\PSC{}} and \mbox{\PCGI{}}, which use locks to ensure consistency in the data.

\subsection{Algorithms that are not lock-free}

Among all algorithms, two are fundamentally different to the rest. These two algorithms are \mbox{\PSC{}}~\cite{graefe2012concurrency,graefe2014transactional} and \mbox{\PCGI{}}, and the difference lies in that they both use locks. As a first reference, Figure~\ref{figs:psc8} shows how the eight-threaded version of \mbox{\PSC{}} utilizes the memory bandwidth of the system. The conclusion is that memory bandwidth is again not hindering the scalability of the algorithms. The real bottleneck is the time \emph{each} thread has to wait while another thread is blocking the resources. Using Vtune we can obtain actual numbers about this. We again take the eight-threaded version of \mbox{\PSC{}} as a reference. Table~\ref{table:waits} shows the total accumulated waiting time of \mbox{\PSC{}} among all threads. Adding up those numbers and dividing by the number of working threads (eight in this case) we obtain that the real waiting time \mbox{\PSC{}} incurs in is $2.105$ seconds on average. Observe in Figure~\ref{figs:psc8} that \mbox{\PSC{}} starts achieving its full potential after the \mbox{$3$-seconds} mark. That is, from the reported total execution time (7.3 secs) of the eight-threaded version of \mbox{\PSC{}} shown in Figure~\ref{figs:psc8}, more than one-third is just waiting time. 

\begin{table}[!htb]
	\small
	\begin{center}
		\begin{tabular}{|c||c|}
		\hline
		{\tt Mutex} & {\tt Wait time (s)}\\
		\hline
		\hline
		Piece lock & $11.671$\\
		Cracker index lock & $5.169$\\
		\hline
		Total & $16.84$\\
		\hline
		Average (Total by $8$) & $2.105$\\
		\hline
		\end{tabular}
	\end{center}
	\caption{Waiting times incurred by the eight-threaded version of \mbox{\PSC{}}. Adding up and dividing by the number of threads we obtain that $2.105$ seconds out of the total execution time of \mbox{\PSC{}} are waiting time.}
	\label{table:waits}
\end{table}

As a quick reference, \mbox{\PCGI{}} builds a range-partitioning before \mbox{\PSC{}} kicks in. This range partitioning alleviates the waiting time, as now a thread does not have to block the whole column when performing the very first crack. As given by Vtune, the total waiting time of \mbox{\PCGI{}} drops to about $1$ second, from the $2.1$ secs of \mbox{\PSC{}}. Thus, the range-partitioning cuts in half the waiting times.

The correctness of \mbox{\PSC{}} and \mbox{\PCGI{}} depends on the use of locks. Therefore, waiting times will always hinder the scalability of these algorithms.

\subsection{Sorting algorithms}

Let us again for the sake of brevity focus only on the eight-threaded versions of \mbox{\PRS{}} and \mbox{\PRPRS{}}. Observing the times shown in Table~\ref{table:sort-comp}, we see that when it comes to \emph{pure} sorting, \mbox{\PRPRS{}} is about 100 milliseconds faster than \mbox{\PRS{}} --- the total execution time of \mbox{\PRPRS{}} is also about 100 milliseconds shorter than that of \mbox{\PRS{}}. We would like to point out that the query code in both algorithms is the same (binary search), and both algorithms only differ in the way they sort the data. How both algorithms utilize the memory bandwidth of the system is shown in Figures~\ref{figs:prs8} and~\ref{figs:prprs8} for \mbox{\PRS{}} and \mbox{\PRPRS{}} respectively. From those figures we can clearly see that stage (\num{3}), \emph{i.e.}, copying the data to the final space where the sorted data will be found, is an ``extra'' part that \mbox{\PRS{}} does in comparison to \mbox{\PRPRS{}}. This explains the lower speedups of \mbox{\PRS{}}. From Table~\ref{table:sort-comp} we can see that the sorting stage of \mbox{\PRPRS{}} scales essentially linearly w.r.t.~the number of working threads. So we will pass onto discussing the speedups achieved by the query part of the algorithms. Since the query code of \mbox{\PRS{}} and \mbox{\PRPRS{}} is the same, we will argue over \mbox{\PRPRS{}} only.

As we already mentioned, the query part of \emph{all} sorting algorithms performs a binary search to filter the elements that belong to the result, and then it (sequentially) aggregates over all those elements. That is, there are no hidden overheads in the code executing the queries. The poor scaling of the query part is due to NUMA effects. In contrast to the chunked algorithms, the sorted column in \mbox{\PRS{}} and \mbox{\PRPRS{}} is shared by both NUMA regions. Queries are served as they come, and to answer a query, this query is assigned to a free thread. This thread then jumps to the NUMA region the query belongs to and filters and aggregates over the corresponding elements. Using Vtune we can quantify the cache misses that are served by the same NUMA region, and the cache misses that are served by a different NUMA region --- and thus also being served slower. For the former, as reported by Vtune, we obtain $175,520,000$ cache misses on average (over ten runs of \mbox{\PRPRS{}}). For the latter we obtain $197,700,000$ on average. So there are $12\%$ more remote cache misses as local cache misses, and overall $53\%$ of the total number of caches misses are remote. This strongly contrast against the chunked algorithms for example, where the number of remote cache misses is negligible, roughly $1\%$ of the total. 

The only way we could get rid of such a high number of NUMA effects is by designing \mbox{\PCRS{}}. There, we sacrifice the fully sorted column --- although, if later needed, the sorted chunks can be merged using NUMA-aware merge procedures like the ones discussed in~\cite{balkesen2013multi}. This sacrifice, nevertheless, comes with the speedups our system allows, as we already discussed.

Having presented the core part of our experiments, we now turn to present other experiments in which we vary certain parameters of the workload such as input size, tuple configuration, query access pattern, distribution of input data, and query selectivity. For all following experiments we tested only the $8$-threaded versions of the algorithms.

\section{Additional experiments}\label{sec:additional}

\subsection{Scaling input size by a factor of ten}\label{sec:experiments:input}

We tested the algorithms with an input size of one billion, \emph{i.e.}, ten times larger than the input size of the previous experiments. The scalability of the algorithms is shown in Table~\ref{table:scalability}. Times are again given in seconds. The shown factors are w.r.t.~the times shown in Tables~\ref{table:sc-comp} to~\ref{table:sort-comp} for the $8$-threaded algorithms. This time nonetheless, all shown factors represent slowdowns due to scaling of the input size. Other than input size, the workload is as described in~\S~\ref{sec:intro:workload}.

\begin{table}[!htb]
	\footnotesize
	\begin{center}
		\begin{tabular}{l c c c c c c}
		\hline
		\algorithm & $\benchb$ & \slowdown & $\bencha$ & \slowdown\\
		& {\tt time} & & {\tt time} &\\
		\hline
		\hline
		\PSC{} & $5.328$ & $9.087\times$ & $59.69$ & $10.02\times$\\
		\PCSC{} & \cellcolor{color}$1.156$ & $9.816\times$ & $31.71$ & $9.482\times$\\
		\PCGI{} & $5.088$ & $12.62\times$ & $40.4$ & $9.298\times$\\
		\PCCGI{} & $3.414$ & $9.934\times$ & \cellcolor{color}$26.64$ & $9.29\times$\\
		\PRS{} & $12.31$ & $13.59\times$ & $45.72$ & $10.77\times$\\
		\PRPRS{} & $9.093$ & $11.59\times$ & $42.75$ & $10.27\times$\\
		\PCRS{} & $7.418$ & $9.307\times$ & $27.43$ & $9.329\times$\\
		\end{tabular}
		\caption{Scale factors of the algorithms when increasing input size by a factor of ten.}
		\label{table:scalability}
	\end{center}
\end{table}

We observe that all algorithms scale gracefully as the input size increases. The slowdown is essentially linear, \emph{i.e.}, we observe times that are essentially ten times slower, although sorting algorithms suffer the most. 

\subsection{Different tuple configuration}\label{sec:experiments:config}

Until now we have assumed that the entries in the cracker column $B$ are pairs $(key, rowID)$ of $4$-byte positive integers each. While this assumption seems in general reasonable, it could also be limiting in some cases. Thus, here we show how the algorithms perform when used with pairs $(key, rowID)$ of $8$-byte positive integers generated uniformly at random over the interval $\left[0, 2^{64}\right)$. The results of these experiments can be seen in Table~\ref{table:tuple}. Times are shown in seconds, and the shown (slowdown) factors are w.r.t.~the times shown in Tables~\ref{table:sc-comp} to~\ref{table:sort-comp} for the $8$-threaded algorithms. 

Other than tuple configuration, the workload is as described in~\S~\ref{sec:intro:workload}. That is, there are $100$ million entries, and $10,000$ queries with selectivity $1\%$ are performed. Queries are this time, of course, $8$-byte pairs of positive integers generated uniformly at random (respecting selectivity). For sorting, we considered the numbers still as radix-$2^{8}$ numbers, \emph{i.e.}, radix sort requires now twice as many passes to fully sort the numbers, \emph{i.e.}, eight passes.

\begin{table}[!htb]
	\footnotesize
	\begin{center}
		\begin{tabular}{l c c c c}
		\hline
		\algorithm & $\benchb$ & \slowdown & $\bencha$ & \slowdown\\
		& {\tt time} & & {\tt time} &\\
		\hline
		\hline
		\PSC{} & $0.7122$ & $1.215\times$ & $9.887$ & $1.66\times$\\
		\PCSC{} & \cellcolor{color}$0.2032$ & $1.726\times$ & $5.426$ & $1.622\times$ \\
		\PCGI{} & $0.6675$ & $1.656\times$ & $7.958$ & $1.831\times$\\
		\PCCGI{} & $0.5841$  & $1.7\times$ & \cellcolor{color}$4.914$ & $1.714\times$\\
		\PRS{} & $1.549$ & $1.709\times$ & $8.16$  & $1.922\times$\\
		\PRPRS{} & $1.188$ & $1.514\times$ & $7.859$ & $1.888\times$\\
		\PCRS{} & $1.078$ & $1.352\times$ & $5.117$ & $1.74\times$\\
		\end{tabular}
		\caption{Scale factors of the algorithms when increasing entry size from $4$-byte pairs to $8$-byte pairs.}
		\label{table:tuple}
	\end{center}
\end{table}

As we can see, there is a generalized two-fold slowdown in all algorithms w.r.t.~total execution time; as expected for a two-fold increment in entry size. In this regard, \PCSC{} is being affected the least, and \PRS{} being affected the most. Initial response time seems to scale more gracefully for all algorithms. With respect to the shown slowdowns of the sorting algorithms we would like to point out that they perform much more work than all other algorithms. Sorting algorithms have to not only shuffle data that is twice as big, which is where the slowdown of all other algorithms stems from, but also they have to do twice as many passes to sort the numbers. Thus, we think that the scale factors of sorting algorithms should be considered exceptionally good.

\subsection{Skewed query access pattern}\label{sec:experiments:skewness}

So far we have seen how the algorithms scale w.r.t.~to input size and different tuple configurations. Now, we show how the algorithms perform when the queries are skewed, \emph{i.e.}, they are no longer uniformly distributed over the range of values the input keys fall into. For these experiments we have generated the queries from a normal distribution with mean $2^{31}$ and standard deviation $2^{28}$. That is, queries are now tightly concentrated around $2^{31}$. 

 As originally stated, we generated $10,000$ queries with selectivity $1\%$, and run the algorithms over $100$ million entries of $4$-byte positive integers generated uniformly at random. The results of the experiments can be seen in Table~\ref{table:skewness}. Times are shown in seconds, and the shown factors are w.r.t.~the times shown in Tables~\ref{table:sc-comp} to~\ref{table:sort-comp} for the $8$-threaded algorithms. A factor larger than one represents a slowdown and a factor smaller than one represents a speedup.

\begin{table}[!htb]
	\footnotesize
	\begin{center}
		\begin{tabular}{l c c c c}
		\hline
		\algorithm & $\benchb$ & \factor & $\bencha$ & \factor\\
		& {\tt time} & & {\tt time} &\\
		\hline
		\hline
		\PSC{} & $0.7489$ & $1.277\times$ & $6.928$ & $1.163\times$\\
		\PCSC{} & \cellcolor{color}$0.1286$ & $1.092\times$ & $2.81$ & $0.8402\times$\\
		\PCGI{} & $0.4317$ & $1.071\times$ & $6.7$ & $1.542\times$\\
		\PCCGI{} & $0.342$ & $0.9953\times$ & \cellcolor{color}$2.609$ & $0.9101\times$\\
		\PRS{} & $0.9038$ & $0.9972\times$ & $4.242$ & $0.9993\times$\\
		\PRPRS{} & $0.7852$ & $1.001\times$ & $4.161$ & $0.9996\times$\\
		\PCRS{} & $0.7989$ & $1.002\times$ & $2.85$ & $0.9694\times$\\
		\end{tabular}
		\caption{Scale factors of the algorithms when queries are skewed.}
		\label{table:skewness}
	\end{center}
\end{table}

We observe that, w.r.t.~initial response time, \PSC{} was affected the most. The explanation for that is the following. As now the queries are tightly concentrated around a certain area, most threads try to access the same area, but this incurs into a great deal of locking. One thread locks the region it partitions while the others are forced to wait. Once done, this thread now has to aggregate over a continuous region, but this region is most probably being locked by another thread partitioning it. So, as it seems, the very first query takes longer to finish just because more threads are accessing the same region. This is the effect of skewness in the query access pattern. \PCGI{} is also being hit due to the same argument, since it uses \PSC{} after range-partitioning the input. Moreover, as \emph{all} queries are tightly concentrated around the same value, there is essentially no difference between \PSC{} and \PCCGI{}, as they lock the same areas. This of course affects \PCCGI{} more than \PSC{}. All other algorithms are completely oblivious to query access patterns. Thus, the shown factors for \PCSC{}, \PCCGI{}, \PRS{}, and \PRPRS{} are the simply variations obtained from different measurements. The observations stated in~\S~\ref{sec:experiments:summarizing} still hold.

\subsection{Skewed input data}\label{sec:experiments:skewedInput}

Here we show how the algorithms perform when the input data is also skewed. Until now we have assumed that the input data is uniformly distributed over a certain range. While this is usually a fair assumption, as it gives a general idea of the performance of an algorithm, in reality, input data might be biased. For these experiments we have chosen to draw the input elements, $4$-byte long, at random from a truncated normal distribution over $\left[0, 2^{32}\right)$. This distribution has mean $2^{31}$, and standard deviation $\left\lceil 2^{31}/3\right\rceil$. With these parameters, the whole interval $\left[0, 2^{32}\right)$ is three standard deviations from the mean; where roughly $68\%$ of the input keys fall in the interval $\left[2^{31} - \left\lceil 2^{31}/3\right\rceil, 2^{31} + \left\lceil 2^{31}/3\right\rceil\right]$, as opposed to the roughly $33\%$ we could expect from the perfect uniform distribution. Other than the distribution of the input keys, the workload is exactly as described in~\S~\ref{sec:intro:workload}. 

The results of this experiments can be seen in Table~\ref{table:skewedInput}. Times are shown in seconds, and the shown factors are w.r.t.~the times shown in Tables~\ref{table:sc-comp} to~\ref{table:sort-comp} for the $8$-threaded algorithms. A factor larger than one represents a slowdown and a factor smaller than one represents a speedup.

\begin{table}[!htb]
	\footnotesize
	\begin{center}
		\begin{tabular}{l c c c c}
		\hline
		\algorithm & $\benchb$ & \factor & $\bencha$ & \factor\\
		& {\tt time} & & {\tt time} &\\
		\hline
		\hline
		\PSC{} & $0.4184$ & $0.7136\times$ & $6.179$ & $1.037\times$\\
		\PCSC{} & \cellcolor{color}$0.1071$ & $0.9093\times$ & $3.354$ & $1.003\times$\\
		\PCGI{} & $0.3874$ & $0.9609\times$ & $4.353$ & $1.002\times$\\
		\PCCGI{} & $0.3363$ & $0.9787\times$ & \cellcolor{color}$2.871$ & $1.001\times$\\
		\PRS{} & $1.465$ & $1.616\times$ & $4.831$ & $1.138\times$\\
		\PRPRS{} & $1.226$ & $1.562\times$ & $4.623$ & $1.11\times$\\
		\PCRS{} & $0.7535$ & $0.9454\times$ & $2.882$ & $0.9801\times$\\
		\end{tabular}
		\caption{Scale factors of the algorithms when the input data is not uniformly distributed.}
		\label{table:skewedInput}
	\end{center}
\end{table}

From the experiments we can see that \PSC{} is the algorithm that benefited the most w.r.t.~initial response time, while \mbox{\PRS{}} is the algorithm that benefits the least. Sorting algorithms were obviously expected to benefit the least, as the skewness of the input breaks the balance in which the work is split among threads. Yet, it is very interesting to observe how radix-sorting methods, in particular \mbox{\PCRS{}}, keep being very competitive.

\subsection{Varying selectivity}\label{sec:experiments:selectivity}

\begin{figure}[!htb]
	\begin{center}
		\includegraphics[scale=0.7]{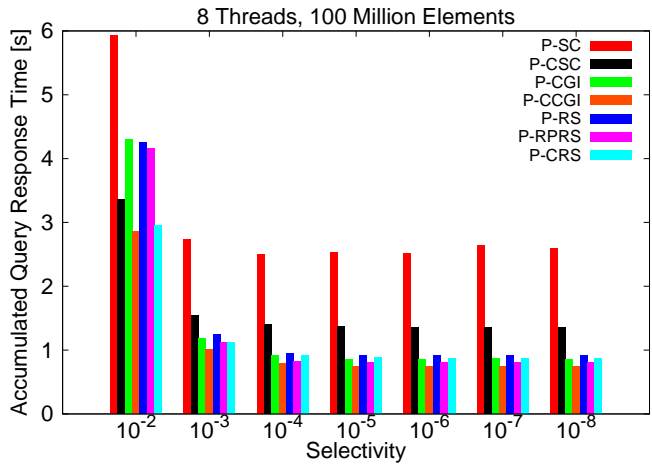}
	\end{center}
	\caption{Effect of varying selectivity from $10^{-2}$ ($1\%$ of the data) to $10^{-8}$ (point query). Ten thousand queries are performed each time.}
	\label{figs:selectivity}
\end{figure}

So far, all experiments considered a selectivity of $1\%$, as this is what is usually used in the literature. Nevertheless, a selectivity of $1\%$ can be deemed to be too low for actually using an index structure at all. Also, the querying part of a query with a low selectivity might overshadow the actual index access. Thus, in this section, we present the results of an experiment where we vary the selectivity from the usual $1\%$ down to highly selective point queries in logarithmic steps. The total number of queries stays at $10,000$ each time. In Figure \ref{figs:selectivity} we show the accumulated query response times for the individual methods under the variation of the selectivity.

Several observations can be made in Figure \ref{figs:selectivity}. First of all, a selectivity higher than $10^{-4}$ ($0.01\%$) does not affect the overall runtime anymore, as the querying part becomes negligible. At that point, the runtime of all methods is determined only by the index creation/maintenance. We can also see that for higher selectivities, the relative order of the methods in terms of performance changes. For a selectivity of $10^{-2}$, \PCRS{} performs much better than \PRS{} and \PRPRS{}. This is no longer the case for higher selectivities, in fact, \PRPRS{} suddenly performs the best of all three from $10^{-3}$ on. The reason for this is that the advantage of \PCRS{} lies in the querying part, which can be performed locally in the chunks due to NUMA-awareness, while \PRS{} and \PRPRS{} suffer from remote accesses. When selectivity increases, index access overshadows aggregation costs. In \mbox{\PCRS{}} all threads work towards answering \emph{every} query, while in the other methods \emph{every} thread answers a different query. That is, the former performs $8$ times more index accesses than the latter. Furthermore, we can also observe that \PCGI{} benefits from high selectivities. From $10^{-3}$ on, its performance is very close to that of the best remaining methods. This improvement results from the fact that from $10^{-3}$ on, a query fits into a partition of the range-partitioning. That is, a thread must lock, in the beginning, at most two partitions of the range-partitioning. The likelihood that any two threads require the same partition is very small. Thus, \PCGI{} becomes mostly lock-free. Overall, we can see that \PCCGI{} shows the best accumulated query response time for all tested selectivities, albeit negligible differences for high-selectivity queries.

\section{Lessons Learned and Conclusion}\label{sec:conclusions}\label{sec:experiments:summarizing}

The most important lessons learned from our initial set of experiments are the following: 
\begin{enumerate}
\item[\num{1}.] Parallel-chunked standard cracking (\PCSC{}) is indisputably the fastest algorithm when answering the very first query (initial response time). Nevertheless, this algorithm is easily outperformed w.r.t.~total time by other algorithms, such as \PCCGI{} and \PCRS{}, as the number of threads increases. 
\item[\num{2}.] Parallel-chunked coarse-granular index (\PCCGI{}) is indisputably the fastest algorithm w.r.t.~accumulated query response time. Moreover, we can observe the following from the core part of our experiments in \S~\ref{sec:experiments}: While \PCSC{} has the best initial response time, it can perform only roughly ten queries while \PCCGI{} is building its range partitioning, see Figure~\ref{figs:8-threaded}. From that point on, \mbox{\PCCGI{}} is already faster than \mbox{\PCSC{}}. Finally, we could also see that \PCCGI{} is highly oblivious to all parameters of the workload discussed by us, which include input size, tuple configuration, query access pattern, distribution of input data, and query selectivity. Thus, \mbox{\PCCGI{}} seems to be the algorithm to choose in practice.
\item[\num{3}.] To our very own surprise, parallel range-partitioned radix sort (\mbox{\PRPRS{}}) turned out to be faster than the previously-known parallel radix sort \PRS{}, while also requiring significantly less extra space. In particular, and of independent interest, we can observe that \mbox{\PRPRS{}} scales essentially linearly w.r.t.~the number of threads when sorting. From our experiments and profiling we can expect that \PRPRS{} will scale more gracefully than the (parallel) adaptive indexing techniques herein considered. This in turn narrows the gap between sorting algorithms and adaptive indexing techniques even further.
\item[\num{4}.] Yet more interesting was to observe how fast \PCRS{} (also sorting-based) earns itself a place among the best algorithms as the number of threads increases, its initial response time is comparable to that of full sorting, but it catches up very quickly, as the query part is very efficient. In the end it is among the fastest algorithms.
\item[\num{5}.] From all our experiments we could observe that our algorithms perform better than existing solutions --- where the chunked algorithms \PCSC{}, \PCCGI{}, and \PCRS{} stand out as the best algorithms overall.
\item[\num{6}.]  In the end, and for each column of a table, a query optimizer has to make a decision whether to not index, to adaptively index or to create a full index. 
Already for the single-threaded algorithms evaluated in~\cite{schuhknecht2013uncracked} we learned that the tipping point may be only after 10~queries, \emph{i.e.},~see Figure~\ref{figs:single-threaded} and observe that \CGI{} is faster than \SC{} already after 10~queries, \RS{} is faster than \SC{} in less than 1,000~queries. Similar observations hold for the parallel case, \emph{i.e.},~see Figure~\ref{figs:8-threaded}. However, we also observe that the tipping point between our fastest sorting-based method, \PCRS{}, and our fastest adaptive indexing method (w.r.t.~initial response), \PCSC{}, moves left to $\approx 500$~queries. As the sorting methods are CPU-bound in their \textit{sorting phases}, see Figure~\ref{figs:8-threadedbw}, we expect the tipping point to move even further left as we add even more threads. 
\item[\num{7}.]  As the overall runtime of the best adaptive indexing and sorting methods, as well as coarse-granular methods, get closer and closer in the parallel case, it is worth remembering that those methods are not exactly comparable. The sorting methods create interesting orders which may further be exploited for query processing, \emph{e.g.},~join processing and aggregations. In addition, the algorithms using a semantically meaningful partition such as \mbox{\PRS{}} and \mbox{\PRPRS{}} create partitions that could be directly exploited to perform NUMA-aware joins (take for instance the popular NUMA-version of the traditional disk-based Grace-Hash Join: rather than partitioning the inputs along the criterium whether they fit into main memory, they partition until the partition pairs fit into a NUMA-region). Thus even though those methods might be slower for a given number of queries, their additional properties may make the overall query faster; hence moving the tipping point even further left.  We are planning to investigate this as part of future work.
\end{enumerate}

\section*{Future Work}

We are currently extending the current work to handle a more realistic workload, as already pointed out in \S~\ref{sec:intro:understanding}. This is way more challenging than in the single-threaded case already investigated in~\cite{schuhknecht2013uncracked}. We will report our findings in a companion paper.

\section*{Acknowledgement}

We would like to thank Felix Halim and Stratos Idreos for kindly providing the source code for \PSC{} used in~\cite{graefe2012concurrency, graefe2014transactional}. The code of our implementations used for our experiments will be made publicly available upon publication of this paper. 

Research partially supported by BMBF.

{\small
\bibliographystyle{abbrv}
\bibliography{dbCracking-MultiCore} 

\begin{thebibliography}{10}

\bibitem{intel-compiler}
Intel\textsuperscript{\textregistered} {C}omposer {X}{E} {S}uites.
\newblock \url{http://software.intel.com/en-us/intel-composer-xe}. Last
  accessed: 25/02/2014.

\bibitem{intel-vtune}
Intel\textsuperscript{\textregistered}
  {V}{T}une\textsuperscript{\texttrademark} {A}mplifier {X}{E} 2013.
\newblock \url{http://software.intel.com/en-us/intel-vtune-amplifier-xe}. Last
  accessed: 25/02/2014.

\bibitem{albutiu2012massively}
M.-C. Albutiu, A.~Kemper, and T.~Neumann.
\newblock Massively parallel sort-merge joins in main memory multi-core
  database systems.
\newblock {\em Proceedings of the VLDB Endowment}, 5(10):1064--1075, 2012.

\bibitem{balkesen2013multi}
C.~Balkesen, G.~Alonso, and M.~Ozsu.
\newblock Multi-core, main-memory joins: Sort vs. hash revisited.
\newblock {\em Proceedings of the VLDB Endowment}, 7(1), 2013.

\bibitem{donald1999art}
E.~K. Donald.
\newblock The art of computer programming.
\newblock {\em Sorting and searching}, 3, 1999.

\bibitem{graefe2012concurrency}
G.~Graefe, F.~Halim, S.~Idreos, H.~Kuno, and S.~Manegold.
\newblock Concurrency control for adaptive indexing.
\newblock {\em Proceedings of the VLDB Endowment}, 5(7):656--667, 2012.

\bibitem{graefe2014transactional}
G.~Graefe, F.~Halim, S.~Idreos, H.~Kuno, S.~Manegold, and B.~Seeger.
\newblock Transactional support for adaptive indexing.
\newblock {\em The VLDB Journal}, pages 1--26, 2014.

\bibitem{graefe2011benchmarking}
G.~Graefe, S.~Idreos, H.~Kuno, and S.~Manegold.
\newblock Benchmarking adaptive indexing.
\newblock In {\em Performance Evaluation, Measurement and Characterization of
  Complex Systems}, pages 169--184. Springer, 2011.

\bibitem{halim2012stochastic}
F.~Halim, S.~Idreos, P.~Karras, and R.~H. Yap.
\newblock Stochastic database cracking: Towards robust adaptive indexing in
  main-memory column-stores.
\newblock {\em Proceedings of the VLDB Endowment}, 5(6):502--513, 2012.

\bibitem{hoare1962quicksort}
C.~A. Hoare.
\newblock Quicksort.
\newblock {\em The Computer Journal}, 5(1):10--16, 1962.

\bibitem{idreos2007database}
S.~Idreos, M.~L. Kersten, and S.~Manegold.
\newblock Database cracking.
\newblock In {\em CIDR}, pages 68--78, 2007.

\bibitem{idreos2007updating}
S.~Idreos, M.~L. Kersten, and S.~Manegold.
\newblock Updating a cracked database.
\newblock In {\em Proceedings of the 2007 ACM SIGMOD international conference
  on Management of data}, pages 413--424. ACM, 2007.

\bibitem{idreos2009self}
S.~Idreos, M.~L. Kersten, and S.~Manegold.
\newblock Self-organizing tuple reconstruction in column-stores.
\newblock In {\em Proceedings of the 2009 ACM SIGMOD International Conference
  on Management of data}, pages 297--308. ACM, 2009.

\bibitem{idreos2011merging}
S.~Idreos, S.~Manegold, H.~Kuno, and G.~Graefe.
\newblock Merging what's cracked, cracking what's merged: adaptive indexing in
  main-memory column-stores.
\newblock {\em Proceedings of the VLDB Endowment}, 4(9):586--597, 2011.

\bibitem{kersten2005cracking}
M.~L. Kersten and S.~Manegold.
\newblock Cracking the database store.
\newblock In {\em CIDR}, pages 213--224, 2005.

\bibitem{kim2009sort}
C.~Kim, T.~Kaldewey, V.~W. Lee, E.~Sedlar, A.~D. Nguyen, N.~Satish,
  J.~Chhugani, A.~Di~Blas, and P.~Dubey.
\newblock Sort vs. hash revisited: fast join implementation on modern
  multi-core cpus.
\newblock {\em Proceedings of the VLDB Endowment}, 2(2):1378--1389, 2009.

\bibitem{lee2002partitioned}
S.-J. Lee, M.~Jeon, D.~Kim, and A.~Sohn.
\newblock Partitioned parallel radix sort.
\newblock {\em Journal of Parallel and Distributed Computing}, 62(4):656--668,
  2002.

\bibitem{leis2013adaptive}
V.~Leis, A.~Kemper, and T.~Neumann.
\newblock The adaptive radix tree: Artful indexing for main-memory databases.
\newblock In {\em ICDE}, 2013.

\bibitem{maus2002arl}
A.~Maus.
\newblock Arl, a faster in-place, cache friendly sorting algorithm.
\newblock {\em Norsk Informatik konferranse NIK}, 2002:85--95, November 2002.

\bibitem{maus2011parl}
A.~Maus.
\newblock A full parallel radix sorting algorithm for multicore processors.
\newblock {\em Norsk Informatik konferranse NIK}, 2011:37--48, November 2011.

\bibitem{Pagh:2001aa}
R.~Pagh and F.~Rodler.
\newblock Cuckoo hashing.
\newblock In {\em ESA 2001}, volume 2161, pages 121--133. Springer Berlin
  Heidelberg, 2001.

\bibitem{rashid2010analyzing}
L.~Rashid, W.~M. Hassanein, and M.~A. Hammad.
\newblock Analyzing and enhancing the parallel sort operation on multithreaded
  architectures.
\newblock {\em The Journal of Supercomputing}, 53(2):293--312, 2010.

\bibitem{schuhknecht2013uncracked}
F.~M. Schuhknecht, A.~Jindal, and J.~Dittrich.
\newblock The uncracked pieces in database cracking.
\newblock {\em Proceedings of the VLDB Endowment}, 7(2), 2013.

\bibitem{stepanov1995standard}
A.~Stepanov and M.~Lee.
\newblock {\em The standard template library}, volume 1501.
\newblock Hewlett Packard Laboratories 1501 Page Mill Road, Palo Alto, CA
  94304, 1995.

\end{thebibliography}
}

\end{document}